# Efficient generation of out-of-plane polarized spin current in polycrystalline heavy metal devices with broken electric symmetries

Qianbiao Liu[1†], Xin Lin[1,2†], Ariel Shaked[3], Zhuyang Nie[4], Guoqiang Yu[4], Lijun Zhu[1,2]*

[1]*State Key Laboratory of Superlattices and Microstructures, Institute of Semiconductors, Chinese Academy of Sciences, Beijing 100083, China*

[2]*College of Materials Science and Opto-Electronic Technology, University of Chinese Academy of Sciences, Beijing 100049, China*

[3]*Cornell University, Ithaca, New York 14850, USA*

[4]*Beijing National Laboratory for Condensed Matter Physics, Institute of Physics, University of Chinese Academy of Sciences, Chinese Academy of Sciences, Beijing 100190, China*

[†] These authors contributed equally

[*] Email: ljzhu@semi.ac.cn

**Spin currents of perpendicularly polarized spins ($z$ spins) by an in-plane charge current have received blooming interest for the potential in energy-efficient spin-orbit torque switching of perpendicular magnetization in the absence of a magnetic field. However, generation of $z$ spins is limited mainly to magnetically or crystallographically low-symmetry single crystals (such as non-colinear antiferromagnets) that are hardly compatible with the integration to semiconductor circuits. Here, we report efficient generation of $z$ spins in sputter-deposited polycrystalline heavy metal devices via a new mechanism of broken electric symmetries in both the transverse and perpendicular directions. Both the dampinglike and fieldlike spin-orbit torques of $z$ spins can be tuned significantly by varying the degree of the electric asymmetries via the length, width, and thickness of devices as well as by varying the type of the heavy metals. We also show that the presence of $z$ spins enables deterministic, nearly-full, external-magnetic-field-free switching of a uniform perpendicularly magnetized FeCoB layer, the core structure of magnetic tunnel junctions, with high coercivity at a low current density. These results establish the first universal, energy-efficient, integration-friendly approach to generate $z$-spin current by electric asymmetry design for dense and low-power spin-torque memory and computing technologies and will stimulate investigation of z-spin currents in various polycrystalline materials.**

## Introduction

Perpendicular magnetization is interesting for the development of spin-orbit torque (SOT)-driven memory and computing technologies with high thermal and magnetic stabilities. However, switching of a stable, uniform perpendicular magnetization by an in-plane current in an energy-efficient and integration-friendly manner has remained a challenge despite the two-decade intensive efforts[1-9]. Instead, a large current and a large in-plane magnetic field along the current direction ($x$ direction) are typically required to switch a uniform perpendicular magnetization by the transverse spins ($y$ spins, polarized in the $y$ direction but flows in the $z$ direction)[10, 11], which limits the energy efficiency and the scalability of the devices. Recently, there has been a blooming interest in generating perpendicular spins ($z$ spins, both polarized and flows in the $z$ direction)[12-23] because of the potential to enable field-free switching of perpendicular magnetization, e.g., via anti-damping and field-like SOTs. So far, discussion of the $z$ spins has been limited mainly to single-crystal films with low crystallographic or magnetic symmetry (e.g., semimetal WTe$_2$ and antiferromagnets Mn$_3$GaN and RuO$_2$)[12,16-23], in which the requirement of single-crystal structures limits the integration to CMOS circuits and thus the technological impact. In contrast, integration-friendly spin-orbit materials, such as polycrystalline heavy metals (HMs), have not been considered as a possible source of the z spins due to their crystallographically and magnetically preserved $\mathcal{M}_{xy}$ and $\mathcal{M}_{xz}$ mirror symmetries about the xy and xz planes (Figs. 1a,b). Note that preservation of any one of the $\mathcal{M}_{xy}$ or $\mathcal{M}_{xz}$ mirror symmetries within the spin-current-generating layer would forbid the imbalanced accumulation of the z spins[24]. Here, we report that the $\mathcal{M}_{xy}$ and $\mathcal{M}_{xz}$ mirror symmetries of polycrystalline HMs, such as sputter-deposited Ta and others on oxidized silicon wafers, can be broken by electric asymmetries induced by device geometries. Consequently, z spins are generated efficiently within the HMs and enable deterministic, nearly full switching of an adjacent perpendicular magnetization without a magnetic field.

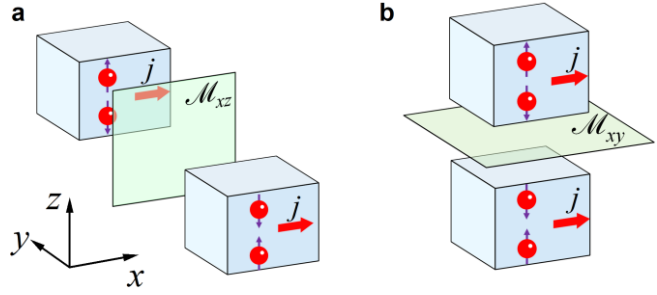

**Fig. 1| $\mathcal{M}_{xy}$ and $\mathcal{M}_{xz}$ mirror symmetries of a heavy metal.** The mirror symmetries (**a**) about the xz plane ($\mathcal{M}_{xz}$) and (**b**) about the xy plane ($\mathcal{M}_{xy}$), each of which can forbid the accumulation of non-equilibrium z spins.

## RESULTS
### Sample details
Samples for this study include Ta (5)/Py (2.1-7.6), Ir (5)/Py (4.6-7.6), Pt (4)/Py (3.3-8.6) bilayers with in-plane magnetic anisotropy, a Ta (5)/FeCoB (2.9) bilayer with in-plane magnetic anisotropy, and a Ta (5)/FeCoB (1.3) bilayer with perpendicular magnetic anisotropy (the numbers in the brackets are layer thicknesses in nanometers, Py = $Ni_{81}Fe_{19}$, FeCoB = $Fe_{60}Co_{20}B_{20}$). Each sample was sputter-deposited on an oxidized silicon substrate and protected by a MgO (1.6)/Ta (1.6) bilayer that is fully oxidized upon exposure to the atmosphere [25]. The samples are then patterned into microstrips and Hall bars by photolithography and ion milling, followed by deposition of Ti (5)/Pt (150) as the contacts for spin-torque ferromagnetic resonance (ST-FMR)[26,27], harmonic Hall voltage (HHV), and current switching measurements.

### Generation and characteristics of z spins
We first characterize the z spins in the Ta (5)/Py (2.7) microstrips (see Fig. 2a,b for an optical microscopy image and a representative ST-FMR spectrum) using in-plane angle-dependent ST-FMR technique subsequently with three-terminal (Fig. 2c) and two-terminal contact configurations (Fig. 2d,e). Considering a magnetic microstrip interacting with a spin current of x-, y-, and z- spin polarizations, the symmetric (S) and anti-symmetric (A) components of the ST-FMR response should vary with the angle ($\varphi$) of the in-plane applied magnetic field relative to the rf current as:

$$S = (S_{DL,y} + S_{SP+heat})\sin2\varphi\cos\varphi + S_{DL,x}\sin2\varphi\sin\varphi + S_{FL+Oe,z}\sin2\varphi \quad (1)$$

$$A = A_{FL+Oe,y}\sin2\varphi\cos\varphi + A_{FL,x}\sin2\varphi\sin\varphi + A_{DL,z}\sin2\varphi, \quad (2)$$

The three terms of Eq. (1) are the sum contributions of the dampinglike SOT of the y spins ($S_{DL,y}$)[26] and the voltages $S_{SP+heat}$ of spin pumping-inverse spin Hall effect of the y spins[28] and the longitudinal spin Seebeck effect and Nernst effects from the resonant heating-induced thermal gradient[29], the dampinglike SOT of the x spins ($S_{DL,x}$)[16], and the sum of the field-like torque of the z spins[16] and the perpendicular Oersted field ($S_{FL+Oe,z}$) [30,31]. Equation (2) includes the transverse effective field from the field-like torque of the $y$ spins [27] and the transverse Oersted field [26], the field-like torque of the $x$ spins [32], and the dampinglike torque of the $z$ spins[12]. According to Eqs. (1) and (2), a nonzero sin2$\varphi$ dependent anti-symmetric ST-FMR signal, i.e., $A_{DL,z}$ can signify the presence of $z$ spins, but a sin2$\varphi$ dependent symmetric ST-FMR signal, i.e., $S_{FL+Oe,z}$ cannot because it can simply from the perpendicular Oersted field [30, 31]. The $S_{SP+heat}$ signal is negligible for all the samples in this work except for Ta/Py samples with the Py thickness greater than 4.6 nm, for the latter $S_{SP+heat}$ is carefully separated from $S_{DL,y}$ (see the Methods).

In Fig. 2c-e, we summarize the magnitudes of the $S$ and $A$ components of the ST-FMR response for the Ta (5)/Py (2.7) as a function of $\varphi$. In the symmetric 3-terminal ST-FMR configuration (Fig. 2c), the fits of the $S$ and $A$ data to Eqs. (1) and (2) reveal the dominance of the $y$ spins contributions ($S_{DL,y}$, $A_{FL+Oe,y}$) and a negligibly small $S_{DL,z}$ and $A_{FL+Oe,z}$ contributions, as is generally true for HM/FM bilayers[30]. Strikingly, significant $A_{DL,z}$ and $S_{FL+Oe,z}$ terms emerge simultaneously when the ST-FMR measurement is performed in a 2-terminal configuration (Fig. 2d,e), suggesting that $z$ spins are generated efficiently by the rf current within the Ta/Py microstrip. In all cases, there is no indication of $x$ spins ($S_{DL,x} = A_{FL,x} = 0$) in this study despite the symmetries (the $\mathcal{M}_{xz}$ mirror symmetries and the rotational symmetry about the $x$ direction, see below) that forbid the accumulation of $z$ spins are all broken, suggesting that having the required symmetry breaking is a necessary but insufficient condition for the generation of spin current.

Characteristically, both $A_{DL,z}$ and $S_{FL+Oe,z}$ require asymmetric current distributions (Fig. 2c vs Fig. 2d,e), reverse sign when the asymmetry of the contact geometry is switched from that in Fig. 2d to the one in Fig. 2e, and strongly dependent on the width ($W$), the length ($L$), the thickness of the FM layer ($t_{Py}$), and the type of the HM of the magnetic microstrip. As shown in Fig. 2f, $A_{DL,z}/S_{DL,y}$ and $S_{FL+Oe,z}/S_{DL,y}$ increase quickly in magnitude as the width increases and relatively slowly as the length decreases. Here, $S_{DL,y}$, which is proportional to the rf current and the efficiency of the dampinglike SOT of the $y$ spins ($\xi^j_{DL,y}$), normalizes $A_{DL,z}$ and $S_{FL+Oe,z}$ to be at the same current density.

We quantify the dampinglike SOT efficiency of $z$ spins ($\xi^j_{DL,z}$) in a HM/FM bilayer using the ST-FMR technique following the relation

$$\xi^j_{DL,z}/\xi^j_{DL,y} = A_{DL,z}/S_{DL,y}\sqrt{1 + 4\pi M_{eff}/H_r}, \quad (3)$$

where $H_r$ and $4\pi M_{eff}$ are the FMR resonance field and the effective demagnetization field of the FM. $\xi^j_{DL,y}$ is -0.011 ± 0.001 for the Ta/Py, 0.020 ± 0.001 for the Ir/Py, and 0.053 ± 0.003 for the Pt/Py as determined from symmetric 3-terminal ST-FMR measurements as a function of the $t_{Py}$ (Method). $\xi^j_{DL,y}$ of ≈-0.011 for the β-Ta/Py is smaller than that of our β-Ta/FeCoB (-0.07) but consistent with the previous reports of Ta/Py samples[33-35]. While the mechanism requires future verification, the different torque appears to agree with the theory [36] that Ta has a positive orbital Hall conductivity and a negative spin Hall conductivity for the $y$ spins and with theory [37] that Ni can more efficiently convert the incident orbital current into spin current due to its stronger spin-orbit coupling than Fe and Co such that Ni can exert a significant orbital Hall SOT that subtracts from the spin Hall SOT on the Py (made mainly of Ni). As we summarize in Fig. 2g, $\xi^j_{DL,z}$ increases monotonically with the thickness of the Py layer. $\xi^j_{DL,z}$ reaches 0.011 for the Ta (5)/Py (7.6), which is already among the highest values of that of single-crystalline antiferromagnet/FM bilayers (0.003-0.019)[17-22]. The $S_{FL+Oe,z}/S_{DL,y}$ data suggests that the field-like torque also increases with the layer thickness of the FM. Both the dampinglike and fieldlike torques are strongly dependent on the type of the HM, i.e., strong for the Ta devices, relatively weak for the Ir and Pt devices (Fig. 2g).

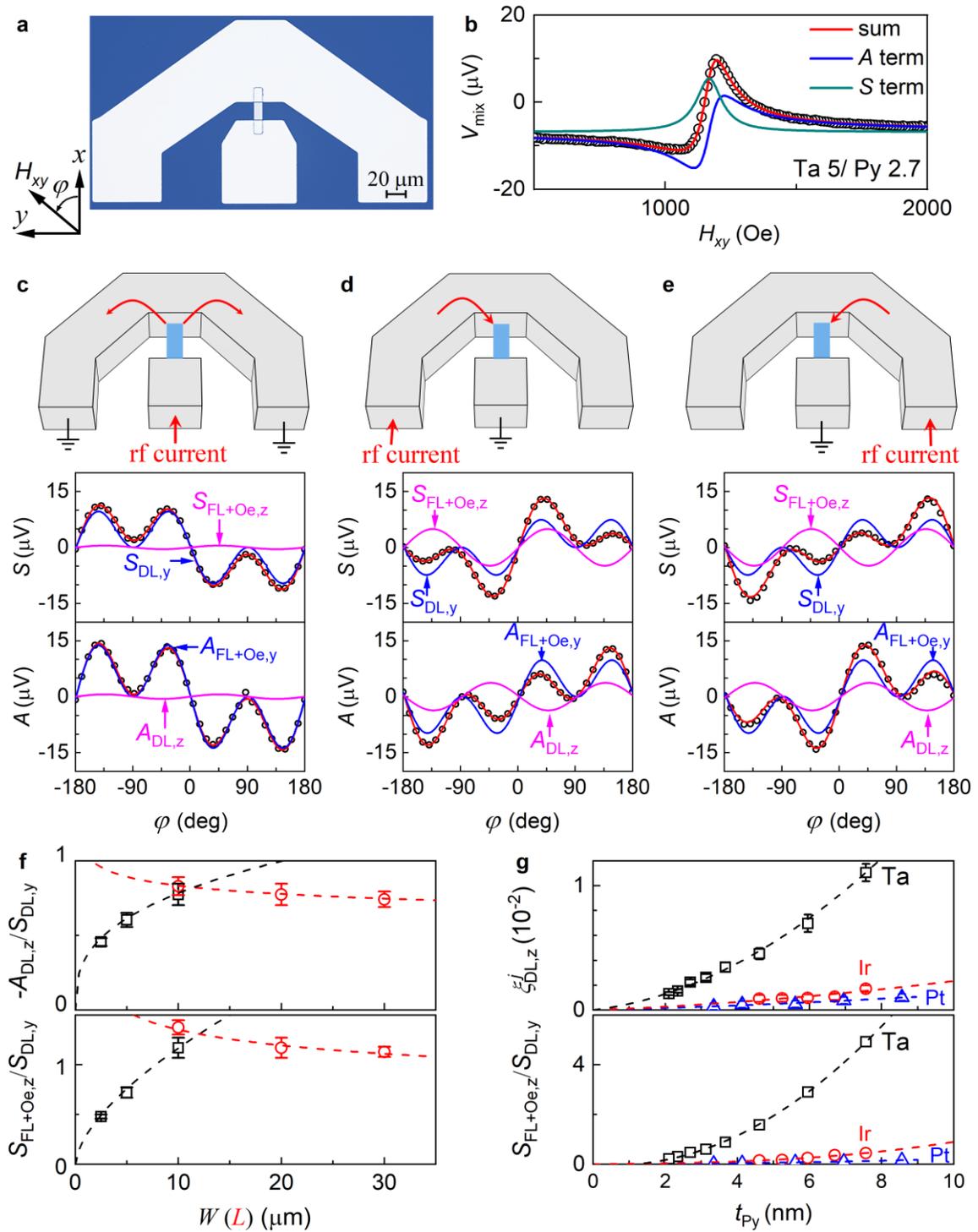

**Fig. 2| ST-FMR characterization of the *z* spins in the Ta/Py strip. (a)** Optical image of a ST-FMR device. **(b)** ST-FMR spectrum for Ta (5)/Py (2.7) (at rf frequency of 8 GHz), which shows a clean symmetric component (*S*, gray) and asymmetric component (*A*, blue). $\varphi$ dependences of *S* and *A* from **(c)** symmetric 3-terminal ST-FMR measurement and **(d** and **e)** asymmetric 2-terminal ST-FMR measurements with the rf current injection from the left and right contact arms, respectively. **(f)** Dependences of $A_{DL,z}/S_{DL,y}$ and $S_{FL,z}/S_{DL,y}$ on the width *W* (*L*=20 μm) and the length *L* (*W*=10 μm) of the Ta (5)/Py (2.7) strips as measured from the 2-terminal ST-FMR configuration in (e). **(g)** Dependences of $\xi_{DL,z}$ and $S_{FL,z}/S_{DL,y}$ on the Py thickness for Ta 5/Py $t_{Py}$ and Pt 4/Py $t_{Py}$ devices (*W*=10 μm, *L*=20 μm) as measured with current injection from the left contract arm. The dashed curves in (f) and (g) guide the eyes.

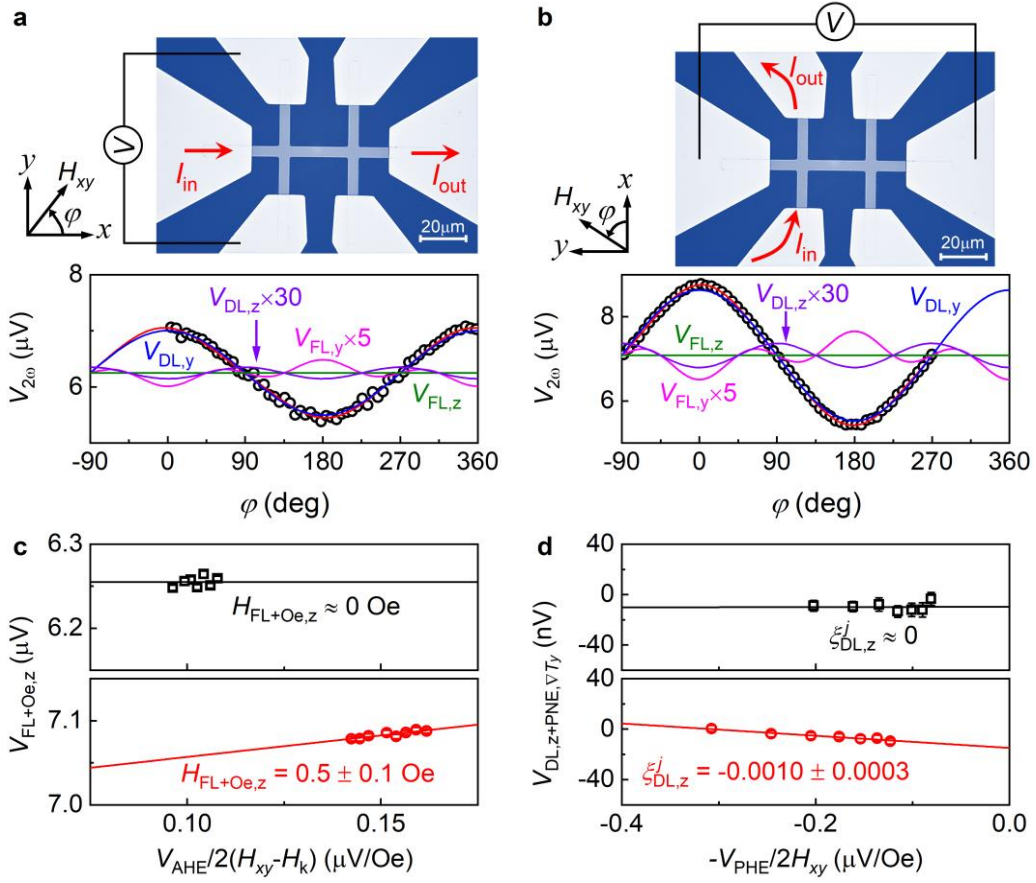

**Fig. 3| HHV characterizations of *z* spins.** $\varphi$ dependence of $V_{2\omega}$ for the Ta (5)/FeCoB (2.9) under an in-plane magnetic field of $H_{xy}$ = 2500 Oe when the current is injected into (**a**) the 5×60 μm² strip from the symmetric contact pair and (**b**) the 5×40 μm² strip from the asymmetric contact pair. In (a) and (b) the solid curves represent different contributions as determined from the best fit of data to Eq. (4); the values of $V_{FL+Oe,y}$ and $V_{DL,z}$ are multiplied by 5 and 30 for clarity. (**c**) $V_{FL+Oe,z}$ vs $V_{AHE}/2(H_{xy}-H_k)$ and (**d**) $V_{DL,z+PNE,\nabla Ty}$ vs $-V_{PHE}/2H_{xy}$ measured with the contact configuration in (a) (back squares) and (b) (red circles).

We further show from HHV measurements that *z* spins can also be generated in polycrystalline Hall bars utilizing electric asymmetries. When a $\varphi$-independent spin current of *y* and *z* spins interacts with an in-plane macrospin, the second HHV is given by

$$V_{2\omega} = V_{DL,y+ANE,\nabla Tz}\cos\varphi + V_{FL+Oe,y}\cos\varphi\cos2\varphi + V_{DL,z+PNE,\nabla Ty}\cos2\varphi + V_{PNE,\nabla Tx}\sin2\varphi + V_{FL+Oe,z}, \quad (4)$$

With

$$V_{DL,y+ANE,\nabla Tz} = V_{AHE}H_{DL,y}/2(H_{xy}-H_k) + V_{ANE,\nabla Tz}, \quad (5)$$

$$V_{FL+Oe,y} = -V_{PHE}H_{FL+Oe,y}/2H_{xy}, \quad (6)$$

$$V_{DL,z+PNE,\nabla Ty} = -V_{PHE}H_{DL,z}/2H_{xy} + V_{PNE,\nabla Ty} \quad (7)$$

$$V_{FL+Oe,z} = V_{AHE}H_{FL+Oe,z}/2(H_{xy}-H_k). \quad (8)$$

Here, $V_{DL,y+ANE,\nabla Tz}$ is the contributions of the dampinglike SOT field of the *y* spins ($H_{DL,y}$)[25] and the anomalous Nernst voltage due to the perpendicular thermal gradient ($V_{ANE,\nabla Tz}$)[38]; $V_{DL,z+PNE,\nabla Ty}$ is the contributions of the dampinglike SOT field of the *z* spins ($H_{DL,z}$)[39,40] and the planar Nernst voltage due to transverse thermal gradient ($V_{PNE,\nabla Ty}$)[41]; $V_{FL+Oe,y}$ is the sum contribution of the fieldlike SOT field of the *y* spins and the transverse Oersted field ($H_{FL+Oe,y}$)[38]; $V_{FL+Oe,z}$ is the contribution of the fieldlike SOT field of the *z* spins[40] and the perpendicular Oersted field [30] ($H_{\text{FL+Oe},z}$); $V_{\text{PNE},\nabla Tx}$ is the planar Nernst voltage due to the longitudinal thermal gradient [42,43]; $H_k$ is the effective perpendicular anisotropic field and estimated to be ≈ -13.7 kOe for the Ta (5)/FeCoB (2.9) from the out-of-plane saturation field; $V_{\text{PHE}}$ and $V_{\text{AHE}}$ are the planar Hall voltage and the anomalous Hall voltage, respectively. Fits of the $V_{2\omega}$ data to Eq. (4) yield the values of $V_{\text{DL},y+\text{ANE},\nabla z}$, $V_{\text{FL+Oe},y}$, $V_{\text{DL},z+\text{PNE},\nabla Ty}$, and $V_{\text{FL+Oe},z}$ for the Ta (5)/FeCoB (2.9) at each magnitude of $H_{xy}$ (Fig. 3a,b). According to Eqs. (7) and (8), the slopes of the linear fits of $V_{\text{DL},z+\text{PNE},\nabla Ty}$ vs $-V_{\text{PHE}}/2H_{xy}$ and $V_{\text{FL+Oe},z}$ vs $V_{\text{AHE}}/2(H_{xy}-H_k)$ give the values of $H_{\text{DL},z}$ and $H_{\text{FL},z+\text{Oe}}$. For the HHV measurements, a sinusoidal electric field $E$ is applied as the excitation using a 3.8 V ac voltage onto the current channel of the Hall cross (60 μm long in Fig. 3a and 40 μm long in Fig. 3b). For the measurement in Fig. 3a, $V_{\text{AHE}} \approx 3.17$ mV, $V_{\text{PHE}} \approx 0.20$ mV, $V_{\text{ANE},\nabla Tz} \approx 0.86$ μV, $V_{\text{PNE},\nabla Tx} \approx 0.024$ μV, $H_{\text{Oe},y} \approx 0.88$ Oe; for the measurement in Fig. 3b, $V_{\text{AHE}} \approx 4.76$ mV, $V_{\text{PHE}} \approx 0.30$ mV, $V_{\text{ANE},\nabla Tz} \approx 1.87$ μV, $V_{\text{PNE},\nabla Tx} = 0.036$ μV, $H_{\text{Oe},y} = 1.32$ Oe. As shown in Fig. 3c,d, $\xi_{\text{DL},z}^j$ and $H_{\text{FL+Oe},z}$ are sizable at the Hall-cross center when current flows in the electrically asymmetric configuration ($H_{\text{FL+Oe},z} \approx 0.5$ Oe) but negligible in the symmetric configuration.

## Generation mechanism of *z* spins

After the generation and characteristics of the *z*-spin current have been established in the ST-FMR and Hall bar devices, we discuss that the *z* spins are generated in the HMs with the $\mathcal{M}_{xy}$ and $\mathcal{M}_{xz}$ mirror symmetries broken by the transverse and perpendicular electric asymmetries of the devices, respectively. The degrees of the transverse and perpendicular electric asymmetries of a device are determined by the device geometry and the material parameters and thus independent of the magnitudes of the injected current or the applied longitudinal electric field $E$. However, the *y* and *z* electric asymmetries of a device can be estimated from their correlation to the corresponding *relative* electric field gradients $\frac{\nabla_y E}{\bar{E}}$ and $\frac{\nabla_z E}{\bar{E}}$ in the measured HM strip when a current is injected into the device ($\bar{E}$ is the volume-average value of $E$ in the HM).

As shown in Fig. 4a, the electric asymmetries and the relative electric field distributions are simulated by finite-element analysis for a HM/Py bilayer device (with the length $L$, the width $W$, the Py thickness $t_{\text{Py}}$, and the HM resistivity $\rho_{xx}$) covered by two contact pads with a dimension of 20μm×40μm×150nm and a resistivity of 24 μΩ cm. In the finite-element analysis, the distribution of charge current density/electric field is calculated following the current continuity equation $\nabla \cdot \bm{j}_c + \frac{\partial \rho}{\partial t} = 0$ and the boundary condition of $\bm{n} \cdot \bm{j}_c = 0$, where $\bm{j}_c = \bm{E}/\rho_{xx}$ is the charge current density, $\rho$ is the charge density, $n$ is the normal direction of the device boundary. The resistivity of the Py is fixed at 47 μΩ cm following our resistivity calibration. The current is modeled as being injected into the left contact along *y* direction and flowing out from the right electric contact along the *x* direction. The element size is set as 200 nm (length) ×200 nm (width) ×0.16 nm (thickness). The finite-element analysis indicates a < 0.5% spatial distribution of $E_x$ relative to the average electric field $\bar{E}$ in the HM due to the electric asymmetries, which can be further quantified by its relative gradients in the longitudinal (*x*), transverse (*y*) and perpendicular (*z*) directions in Fig. 4b. To compare with the results of the ST-FMR experiment (a transport experiment), we plot in Fig. 4c the volume-averaged values of the relative field gradients, i.e., $\frac{\overline{\nabla_x E}}{\bar{E}}$, $\frac{\overline{\nabla_y E}}{\bar{E}}$, and $\frac{\overline{\nabla_z E}}{\bar{E}}$. In contrast to the negligible longitudinal electric asymmetry $\frac{\overline{\nabla_x E}}{\bar{E}}$, the transverse and perpendicular electric asymmetries ($\frac{\overline{\nabla_y E}}{\bar{E}}$ and $\frac{\overline{\nabla_z E}}{\bar{E}}$) that can break the $\mathcal{M}_{xy}$ and $\mathcal{M}_{xz}$ symmetries, respectively, are significant. Since the generation of *z* spins relies on the simultaneous breaking of the $\mathcal{M}_{xy}$ and $\mathcal{M}_{xz}$ symmetries, we further plot the product $-\frac{\overline{\nabla_y E}}{\bar{E}} \frac{\overline{\nabla_z E}}{\bar{E}}$ in Fig. 4d. The magnitude of $-\frac{\overline{\nabla_y E}}{\bar{E}} \frac{\overline{\nabla_z E}}{\bar{E}}$ decreases with $L$ and increases with $W$ and $t$, which agree reasonably with the ST-FMR results of the damping-like and field-like torques in Fig. 2f,g). We note that the simulated dependences of $-\frac{\overline{\nabla_y E}}{\overline{E}}\frac{\overline{\nabla_z E}}{\overline{E}}$ on $L$, $W$, and $t_{Py}$ (Fig. 4d) are slightly stronger than the experimental observation of the torque signals ($A_{DL,z}/S_{DL,y}$, $S_{FL,z}/S_{DL,y}$, $\xi_{DL,z}$ and $S_{FL,z}/S_{DL,y}$ in Fig. 2f,g). This could be attributed to the overlook of the contact resistance between the electrodes and the magnetic strips and the interface resistance at the HM/FM interfaces in the finite-element analysis. It is also possible that the efficiency of the $z$ spin generation is positively correlated to but not exactly a linear function of $-\frac{\overline{\nabla_y E}}{\overline{E}}\frac{\overline{\nabla_z E}}{\overline{E}}$. For HHV measurement with the asymmetric electrodes, the finite-element analysis also suggests a small but sizeable relative electric field gradient $-\frac{\overline{\nabla_y E}}{\overline{E}}\frac{\overline{\nabla_z E}}{\overline{E}} \approx -4.0 \times 10^4$ m$^{-2}$) in the 5×5 μm$^2$ center area of the Ta (5)/FeCoB (2.9) Hall cross. Thus, we conclude that the $\mathcal{M}_{xy}$ and $\mathcal{M}_{xz}$ symmetries (Fig. 1a,b) are broken by the transverse and perpendicular electric asymmetries, enabling the generation of the $z$ spins in the HMs by the longitudinal current induced by $E_x$. It is necessary to note that the absolute values of electric field gradients or $\overline{\nabla_y E \nabla_z E}$ cannot precisely parameterize the degrees of the $\mathcal{M}_{xy}$ and $\mathcal{M}_{xz}$ symmetry breakings. For given degrees of the symmetry breakings (precisely parameterized by $-\frac{\overline{\nabla_y E}}{\overline{E}}\frac{\overline{\nabla_z E}}{\overline{E}}$), the electric field gradients parameterized by $\overline{\nabla_y E \nabla_z E}$ vary with $E^2$. Increase of $E$ increases the electric field gradients but not the degree of the symmetry breakings. Technically, if the $z$ spin Hall conductivity scaled with the absolute values of the field gradients rather than the relative electric field gradients, the ST-FMR voltage signals of the $z$ spins would be out-of-phase second harmonic function of the rf current due to the rectification effect and must not present in the ST-FMR signals we collect from the in-phase first harmonic detection using lock-in amplifier.

Without breaking the $\mathcal{M}_{xy}$ and $\mathcal{M}_{xz}$ symmetries, any electric field $\boldsymbol{E}$, regardless the distribution, cannot generate a $z$-spin current (polarized and flows collinearly in the $z$ direction) via $\boldsymbol{j}_s = \overleftrightarrow{\sigma_{SH}}\boldsymbol{E}$ because the high-symmetry HMs enforce the $z$ spin components of the spin Hall conductivity tensor $\overleftrightarrow{\sigma_{SH}}$ to be zero[40, 44]. Our experimental observation of dampinglike and fieldlike SOTs of the $z$ spins and their close correlation to the degree of the $\mathcal{M}_{xy}$ and $\mathcal{M}_{xz}$ mirror symmetry breakings (as parameterized by $-\frac{\overline{\nabla_y E}}{\overline{E}}\frac{\overline{\nabla_z E}}{\overline{E}}$) strongly suggest that the symmetry of $\overleftrightarrow{\sigma_{SH}}$ of the HMs can be lowered when the $\mathcal{M}_{xy}$ and $\mathcal{M}_{xz}$ mirror symmetries are broken by the electric asymmetries such that the related $z$-spin component of $\overleftrightarrow{\sigma_{SH}}$ becomes sizable. A theory has suggested that the symmetry of the spin Berry curvature can be altered by electric field via the space group[45].

The strong dependences of the dampinglike and field-like SOTs of $z$ spins on the type of the HMs (strong in the Ta/Py, weak in the Ir/Py and the Pt/Py, Fig.2g) cannot be attributed to the difference in the resistivities of the HMs because our simulation has indicated that the degree of the $\mathcal{M}_{xy}$ and $\mathcal{M}_{xz}$ symmetry breakings actually reduces as the HM resistivity increases (Fig. 4d). The latter can be understood as follows. If the dominant source of the current spreading is associated with the resistivity mismatch between the electrodes and the HM/FM channel, the current spreading in the channel should be greatest when the channel resistivity is smallest relative to electrodes. This is easiest to think about in the limiting cases — if the electrode is very resistive compared to the channel then the current will want to flow only through the minimum length path in the electrode and will enter the channel only at the nearest corner of the channel, while if the electrode were superconducting than the interface would be a equipotential and current would enter the channel uniformly around the boundary. Therefore, the $z$-spin component of $\overleftrightarrow{\sigma_{SH}}$ induced by electric asymmetries is sensitive to the type and thus band structure of the HMs, but not simply the resistivity. Electric asymmetries appear to trig the emergence of the $z$-spin component of $\overleftrightarrow{\sigma_{SH}}$ in $\beta$-Ta but less effectively in the Pt and Ir. Theoretical calculations of the spin Hall conductivity tensor and the spin Berry curvature in the presence of broken $\mathcal{M}_{xy}$ and $\mathcal{M}_{xz}$ symmetries would be informative but beyond the scope of our present paper. Note that the strong dependences of the z-spin torques on the contact geometries and layer thicknesses of the devices (Fig. 2; the z-spin torques also increase as the HM thickness increases, not shown) safely exclude any interface effects as the source of the z spins. Any effects (e.g., spin swapping[46]) should be insensitive to the contact geometry of the device and reduce as the layer thicknesses increase, which is contrary to the experimental observation of $\xi_{DL,z}^{j}$ and $S_{FL+Oe,z}/S_{DL,y}$. The strong material dependence of the $S_{FL+Oe,z}/S_{DL,y}$ data in Fig.2g suggest that the perpendicular effective field is predominately from the field-like SOT of z spins rather than from the perpendicular Oersted field [30] related to the transverse gradient of $E$. This is because, for given device dimensions the perpendicular Oersted field, if significant, should be greater for devices with low-resistivity HMs than for those with high-resistivity HM (see $\frac{\nabla_y E}{\bar{E}}$ vs $\rho_{xx}$ in Fig. 4c). The material and thickness dependences of the $S_{FL+Oe,z}/S_{DL,y}$ scale closely with that of $\xi_{DL,z}^{j}$, which is consistent with the torque characteristics of a spin current [47].

We argue that the mechanism of broken electric symmetries we establish in this work is the first proven mechanism that allows to generate z spin currents in polycrystalline samples. In principle, the generation of z spin currents is expected *only* in systems with the simultaneous breaking of the $\mathcal{M}_{xz}$ and $\mathcal{M}_{xy}$ symmetries[24] and can be signified *only* by a nonnegligible *dampinglike* torque of z spins. As we have discussed previously[30], external field-free switching of perpendicular magnetization or the presence of a perpendicular effective magnetic field cannot signify z spins because they can also result from spin-current-unrelated mechanisms, such as perpendicular Oersted field in magnetic structures with spatial non-uniformity or gradients in geometry[30], composition, resistivity, and/or layer thickness, long-range intralayer DMI field in magnetic layers adjacent to a HM or oxide layer[48], or long-range interlayer DMI[49], interlayer exchange coupling, and Néel orange-peel effect[50] in orthogonal FM/HM/FM trilayers, etc.

In the literature, most reports[14,15,51-58] claiming z spin currents only observed field-free switching of perpendicular magnetization and/or a perpendicular effective magnetic field, but no evidence of the dampinglike torque of z spins, leaving the effectiveness of the claimed mechanisms (spin precession or spin swapping) an open question. Particularly, in the polycrystalline bilayers and single layers exhibiting field-free switching[56-58], it remains unclear how the $\mathcal{M}_{xz}$ and $\mathcal{M}_{xy}$ symmetries that forbid the generation of z spins were broken at the same time. Some other reports [40,59-63] claimed a dampinglike torque of z spins from a very small $\sin2\varphi$ dependent antisymmetric ST-FMR signal or a $\cos2\varphi$-dependent second HHV signal by assuming that symmetry breakings by in-plane magnetization generated a z spin current that is independent of the magnetization orientation. However, the generation of z-spin current in a magnetic layer would be allowed by the symmetry argument when the magnetization is parallel, not perpendicular, to the in-plane charge current (see Fig. 3b in Ref. 11 and Fig. 3 and Fig. 4 in Ref. 24). We also find that antisymmetric ST-FMR signals in those works can be fit well without the claimed $\sin2\varphi$ or $\cos2\varphi$ dependent term (see Fig. S1 in the supporting information). We do find two experiments that indeed have provided clear ST-FMR evidence, but no realistic physics origin, of the dampinglike torque of z spins in polycrystalline EuS/Py bilayers[64] and Py single layers[65]. We infer that the z spins in those two works arise likely due to the asymmetric contact (as indicated by the contact misalignment in the optical microscopy image in ref. 66 from the same group) and two-terminal geometry of their ST-FMR devices, and thus can be understood by the mechanism of electric asymmetries we first establish in our present work.

We emphasize that, theoretically, spin filtering at HM/FM interfaces may generate y spins, not z spins[67], while spin rotation[67] and swapping[46] only cause a fieldlike torque, not dampinglike torque, of y spins (not z spins) on the associated FM layer. Experimentally, these interfacial effects have been verified to contribute negligibly small charge-to-spin conversion even when the interfacial spin-orbit coupling is very strong[68]. This is consistent with the absence of any significant z spin current in the HM/FM or FM without electrical asymmetries in this work (Fig. 2a) and the widespread consensus in the past 15 years that no z spins are possible if there was no additional breaking of the $\mathcal{M}_{xy}$ and $\mathcal{M}_{xz}$ mirror symmetries that forbid the accumulation of $z$ spins.[2-20,24]

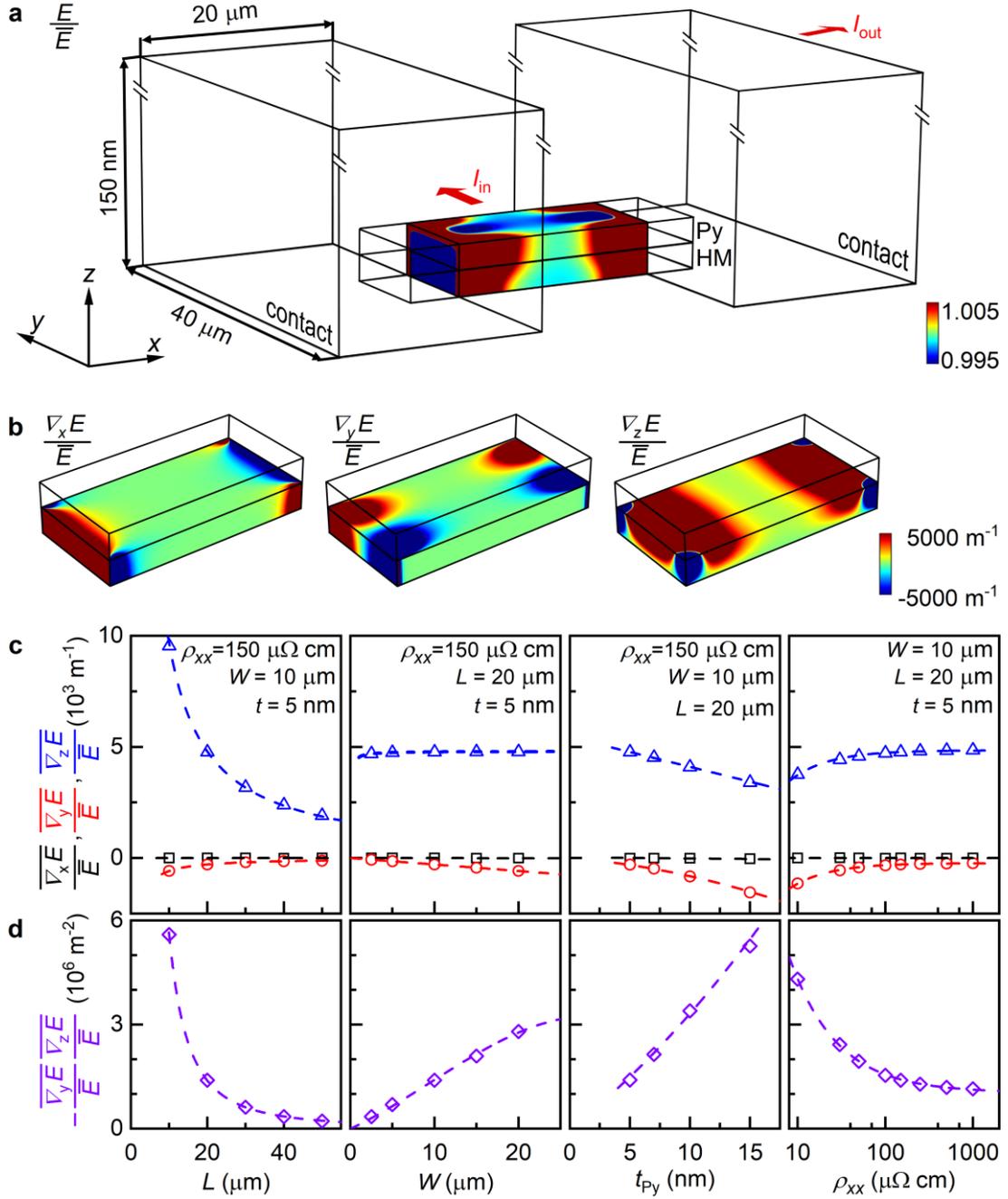

**Fig. 4| Finite-element analysis of the electric asymmetries in the HM/Py ST-FMR device. (a)** Distribution and **(b)** the relative gradients of the longitudinal electric field $E$ ($\frac{\overline{\nabla_x E}}{\overline{E}}$, $\frac{\overline{\nabla_y E}}{\overline{E}}$, and $\frac{\overline{\nabla_z E}}{\overline{E}}$) during 2-terminal ST-FMR measurement in Fig. 2b. The current is injected from the left contact along the $y$ direction and flows out from right pad along the $x$ direction. The results in (a**)** and (b**)** are for the width $W$=10 μm, the length $L$=20 μm, the Py thickness $t_{Py}$=5 nm (rescaled in the plot in (a) for clarity), and the HM resistivity $\rho_{xx}$=150 μΩ cm. **(c)** Average values of the relative gradients of the longitudinal electric field ($\frac{\overline{\nabla_x E}}{\overline{E}}$, $\frac{\overline{\nabla_y E}}{\overline{E}}$, and $\frac{\overline{\nabla_z E}}{\overline{E}}$.) and **(d)** $-\frac{\overline{\nabla_y E}}{\overline{E}} \frac{\overline{\nabla_z E}}{\overline{E}}$ for the ST-FMR devices with different $L$ ($\rho_{xx}$ =150 μΩ cm, $W$=10 μm, $t_{Py}$ =5 nm), $W$ ($\rho_{xx}$ =150 μΩ cm, $L$=20 μm, $t_{Py}$ =5 nm), and $t_{Py}$ ($\rho_{xx}$ =150 μΩ cm, $W$=10 μm, $L$=20 μm), and $\rho_{xx}$ ($W$=10 μm, $L$=20 μm, $t_{Py}$ =5 nm).

**Field-free SOT switching of perpendicular magnetization.**

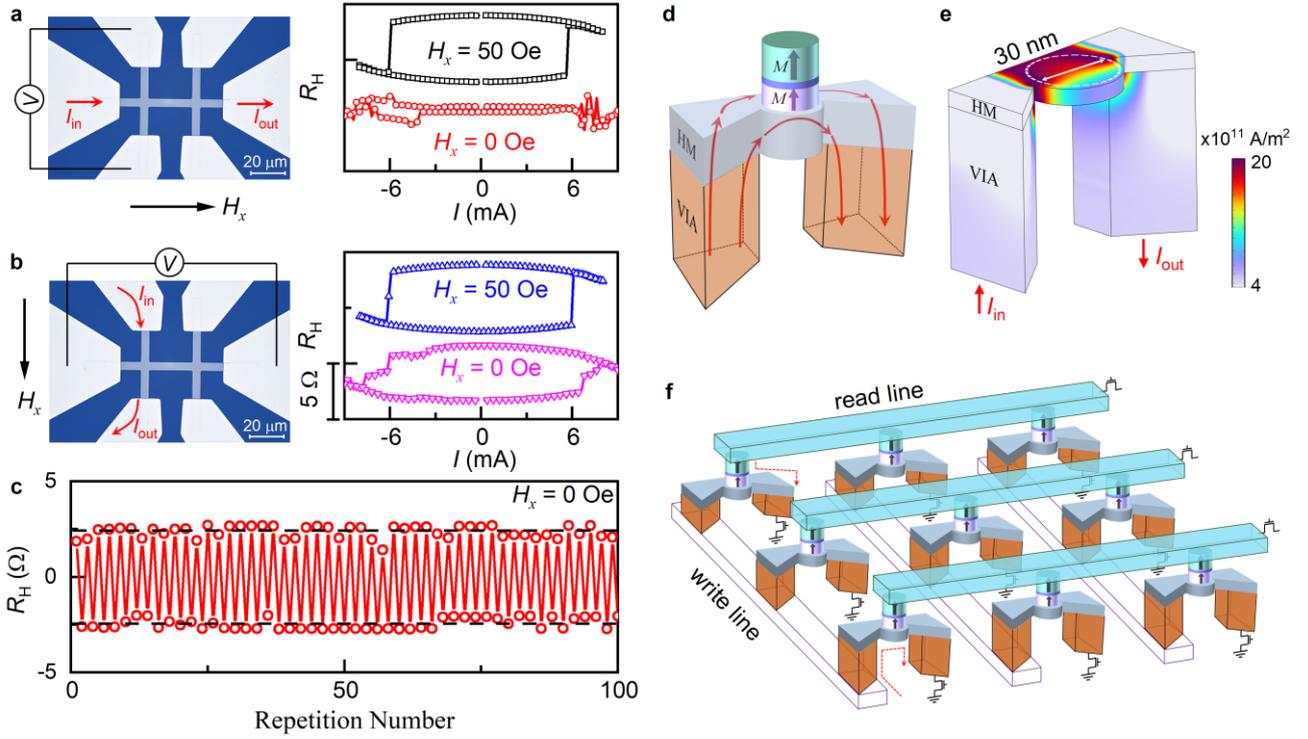

**Fig. 5 Current-driven switching of perpendicular magnetization.** Hall resistance loop of the Ta(5)/FeCoB(1.3) as measured by injecting current via **(a)** symmetric and **(b)** asymmetric contact configurations, under an in-plane field of 50 Oe or 0 Oe along the current direction (defined as the *x* direction). **(c)** Hall resistances collected from the 100 repeats of deterministic magnetization switching by an in-plane current in the absence of a magnetic field. **(d)** Schematic depict of a SOT-MRAM device with enhanced asymmetric geometry design. (**e**) Finite-element analysis result of the current distribution and the relative electric asymmetries in the SOT-MRAM device proposed in (d). **(f)** Integration of asymmetric-geometry-type SOT-MRAM array with electric asymmetries for external-magnetic-field-free operation.

Finally, we demonstrate that the *z* spins induced by electric asymmetries can enable deterministic switching of perpendicular magnetization in the absence of any external magnetic field (Fig. 5a-c). When an in-plane charge current is injected through the asymmetric contact pair (Fig. 5b), the perpendicularly magnetized Ta (5)/FeCoB (1.3) Hall bar (with coercivity of 200 Oe, saturation magnetization of 1300 emu/cm$^3$, perpendicular magnetic anisotropy field of 8200 Oe, and $\xi_{DL,y}^{j}$ of ≈0.07) show a nearly full switching in the absence of an in-plane assisting magnetic field, with the anomalous Hall resistance variation being over 82% of that of the complete switching under a 50 Oe assisting field along the current direction. This ratio is significantly higher than the highest achieved by in-plane CoFeB/Ti/perpendicular CoFeB trilayer (27%)[14] or by single crystals of L1$_1$-CuPt (40%) [15], Mn$_3$Sn (50%) [21], and Mn$_3$SnN (60%)[17]. In contrast, the same Hall bar cannot be switched under zero magnetic field when the in-plane charge current is injected through the symmetric contact pair (Fig. 5a), reaffirming that the occurrence of the observed external field-free magnetization switching is closely related to the electric asymmetries or the *z*-spin current. Importantly, the field-free switching enabled by the SOTs of *z* spins is efficient and deterministic. Such field-free switching only requires a switching current of ≈6 mA (or current density of 1.4×10$^7$ A/cm$^2$), which is close to that of the in-plane field assisted anti-damping torque depinning of domain wall mechanism (Fig. 5a,b). The latter is a very efficient switching mechanism compared to the coherent rotation mechanism[69]. Obviously, the different switching results in Fig. 5a,b cannot be attributed to the Oersted field ($H_{Oe,y} \approx 6.8$ Oe, $H_{FL+Oe,z} \approx 3.0$ Oe at 6 mA) or device shape anisotropy (negligible for the micrometer-scale devices[70]). As shown in Fig. 5c, such external-field-free switching of perpendicular magnetization via electric asymmetries is also highly deterministic.

More generally, the strategy of the generation of unconventional spin current by electric symmetry breakings is integration-friendly and thus technologically promising. As we schematically show in Fig. 5d and simulated using finite-element analysis in Fig. 5e, strong electric asymmetries can be introduced compactly in a short heavy-metal current channel of a nanoscale magnetic tunnel junction, by asymmetric geometry design of the channel and vias of the write current. Such asymmetric contact-geometry magnetic tunnel junctions can be further integrated into SOT-MRAM arrays that can be operated without an external magnetic field (Fig. 5f). Note that this field-free-switching strategy based on $z$-spin current is distinct from non-uniform current-induced perpendicular Oersted field in Pt/FM devices[30] and in-plane spin current non-uniformity in W/CoFeB devices[71]. Generation of $z$-spin current was quantified to be negligible in the Pt/FM devices of Ref. 30, while Ref. 71 made no discussion on $z$ spins or quantification of any SOTs. The device geometry we propose in Fig. 5d is also much more compact and thus integration-friendly than "L"- or "Z"-shaped ones in Refs. 30 and 71.

## Conclusion

We have proposed and demonstrated the efficient generation of the $z$ spin current via electric asymmetries in sputter-deposited polycrystalline heavy metals that are integration-friendly and thus highly preferred for technologies compared to the single crystals of magnetically low-symmetry antiferromagnets or crystallographically non-centrosymmetric semimetals [12-21,72]. The dampinglike spin-orbit torque of the $z$ spins is tuned strongly by varying electric asymmetries of the devices via the length, width, thickness, and by varying the type of the HM of devices. For instance, in Ta/Py devices we observe a strong dampinglike SOT of z spins ($\xi_{DL,z}^j = 0.011$) that is comparable to that of y spins ($|\xi_{DL,z}^j / \xi_{DL,y}^j| = 1$). We demonstrate that the perpendicularly magnetized Ta/FeCoB Hall bar, the core element of perpendicular SOT-MRAMs, can be switched reliably at low current densities due to the presence of $z$ spins, in the absence of an external magnetic field. These results pave the first universal, integration-friendly way to efficiently generate perpendicularly polarized spin current in spin-orbit materials by asymmetric contact design for dense and low-power spin-torque technologies. We also believe these striking results will stimulate the investigation of z-spin currents in various polycrystalline materials.

## Methods:

**Sample preparation:** Magnetic heterostructures are sputter-deposited on oxidized Si substrates. Each sample is protected from oxidization by a MgO (1.6)/Ta (1.6) bilayer that is fully oxidized upon exposure to the atmosphere. No thermal annealing was performed on the samples. The layer thicknesses are estimated using the calibrated deposition rates and the deposition time and then verified by scanning transmission electron microscopy measurements. The base pressure during deposition is below $5\times10^{-9}$ Torr. The argon pressure during deposition is 2 mTorr. The sputtering power is 100 W for MgO but 30 W for all the other materials. These samples are patterned into $5\times60$ μm$^2$ Hall bars by photolithography and ion milling, followed by the deposition of Ti (5)/Pt (150) as the electrical contacts for switching measurements.

**Magnetization and low-frequency electrical measurement:** The saturation magnetization of each sample was measured by a standard vibrating sample magnetometer embedded in a Quantum Design physical properties measurement system. Current is sourced into the Hall bars by a Keithley 6221 or by a SR860 lock-in amplifier, and the Hall voltages of the Hall bars are detected by a SR860 lock-in amplifier.

**ST-FMR analysis:**

The efficiencies of damping-like and field-like spin-orbit torques of perpendicular (*z*) and transverse (*y*) spins can be estimated using the measured dampinglike and field-like SOT fields, $H_{DL,y}$, $H_{DL,z}$, $H_{FL,y}$, and $H_{FL,z}$ following

$$\xi_{DL,y}^{j} = (2e/\hbar) H_{DL,y} \mu_0 M_s t_{FM}/ j_{HM} \quad (M1)$$

$$\xi_{DL,z}^{j} = (2e/\hbar) H_{DL,z} \mu_0 M_s t_{FM}/ j_{HM} \quad (M2)$$

$$\xi_{FL,y}^{j} = (2e/\hbar) H_{FL,y} \mu_0 M_s t_{FM}/ j_{HM} \quad (M3)$$

$$\xi_{FL,z}^{j} = (2e/\hbar) H_{FL,z} \mu_0 M_s t_{FM}/ j_{HM} \quad (M4)$$

where *e* is the electron charge, $\hbar$ the reduced Planck constant, $\mu_0$ the vacuum permeability, $M_s$ the magnetization, $t_{FM}$ the thickness of the magnetic layer, and $j_{HM}$ the current density within the spin-current generator. The magnitudes of $H_{DL,y}$, $H_{DL,z}$, $H_{FL,y}$, and $H_{FL,z}$ can be measured using the spin torque ferromagnetic resonance (ST-FMR). During the measurement, an in-plane rf current ($I_{rf}$) is sourced into the microstrip device and excites magnetization precession via SOTs and Oersted field under certain in-plane applied magnetic field *H*. Mixing of precession-induced magnetoresistance change and the rf current leads to a d.c. voltage ($V_{mix}$) across the magnetic microstrip:

$$V_{mix} = S \frac{\Delta H^2}{\Delta H^2+(H-H_r)^2} + A \frac{\Delta H(H-H_r)}{\Delta H^2+(H-H_r)^2} \quad (M5)$$

where $\Delta H$ and $H_r$ are the FMR linewidth and the resonance field; *S* and *A* are the Lorentzian coefficients of the symmetric and anti-symmetric components. To improve the signal-noise ratio, the rf power is modulated sinusoidally and $V_{mix}$ is detected as the in-phase first harmonic response by a SR860 lock-in amplifier.

Considering a magnetic microstrip interacting with a spin current of *x*-, *y*-, and *z*-spins, the symmetric (*S*) and anti-symmetric (*A*) components of the ST-FMR response should vary with the angle ($\varphi$) of the in-plane applied magnetic field relative to the rf current following Equations (1) and (2). The magnitudes of $S_{DL,y}$, $S_{DL,y}$, $A_{FL,y}$, and $A_{DL,z}$ in Equations (1) and (2) correlate to the magnitudes of $H_{DL,y}$, $H_{DL,z}$, $H_{FL,y}$, and $H_{FL,z}$ as

$$S_{DL,y} = I_{rf} C_{MR} H_{DL,y} \quad (M6)$$
$$S_{DL,z} = I_{rf} C_{MR} (\hbar/2e) H_{DL,z} \quad (M7)$$
$$A_{FL,y} = I_{rf} C_{MR} (H_{FL,y}+H_{Oe,y})\sqrt{1+4\pi M_{eff}/H_r}$$
$$= I_{rf} C_{MR}(\frac{\hbar}{2e} j_{HM} \frac{\xi_{FL,y}^{j}}{\mu_0 M_s t_{FM}} + \frac{j_{HM} d_{HM}}{2})\sqrt{1+4\pi M_{eff}/H_r} \quad (M8)$$
$$A_{DL,z} = I_{rf} C_{MR} H_{DL,z} \sqrt{1+4\pi M_{eff}/H_r} \quad (M9)$$

where $C_{MR}$ is the coefficient related the magnetoresistance of the magnetic layer, $H_{Oe,y}$ is the Oersted field exerted onto the magnetic layer by the adjacent layers, $4\pi M_{eff}$ is the effective demagnetization field of the spin current detector. As shown in Fig. 6, $4\pi M_{eff}$ can be determined from the resonance frequency (*f*) dependence of $H_r$ following the Kittel's equation $f = \gamma/2\pi \sqrt{H_r(H_r + 4\pi M_{eff})}$. Combining Eqs. (M1), (M2), (M6), and (M9), we obtain

$$\xi_{DL,z}^{j}/ \xi_{DL,y}^{j} = A_{DL,z}/S_{DL,y} \sqrt{1+4\pi M_{eff}/H_r}. \quad (M10)$$

Below we further discuss the determination of $\xi_{DL,y}$ and $S_{DL,y}$. According to Equation (1), the $\sin 2\varphi \cos\varphi$ component of the symmetric ST-FMR response may include $S_{DL,y}$ and the sum voltage $S_{SP+heat}$ of the spin pumping-inverse spin Hall voltage of the *y* spins and the voltage of longitudinal spin Seebeck effect and Nernst effects from the resonant heating-induced thermal gradient. While it is the more general case $S_{DL,y} \gg S_{SP+heat}$ for thin HM/FM bilayers, $S_{SP+heat}$ can still become non-negligible and mix with $S_{DL,y}$ in the $\varphi$ dependence of the $S$ component of the ST-FMR response in some samples with a thick FM layer, e.g., the Ta (5)/Py with the Py thickness greater than 4.6 nm in this work. If we define the apparent FMR efficiency as

$$\xi_{FMR} \equiv \frac{S_{DL,y}+S_{SP+heat}}{A_{FL,y}} \frac{e\mu_0 M_s t_{FM} t_{HM}}{\hbar} \sqrt{1+4\pi M_{eff}/H_r}, \quad (M11)$$

with

$$\xi_{FMR,y} \equiv \frac{S_{DL,y}}{A_{FL,y}} \frac{e\mu_0 M_s t_{FM} t_{HM}}{\hbar} \sqrt{1+4\pi M_{eff}/H_r}, \quad (M12)$$

$$\xi_{FMR} \equiv \frac{S_{SP+heat}}{A_{FL,y}} \frac{e\mu_0 M_s t_{FM} t_{HM}}{\hbar} \sqrt{1+4\pi M_{eff}/H_r}. \quad (M13)$$

we have

$$\frac{1}{\xi_{FMR,y}} = \frac{1}{\xi_{FMR}-\xi_{FMR,SP+heat}} = \frac{1}{\xi^j_{DL,y}}(1+\frac{\hbar \xi^j_{FL,y}}{e\mu_0 M_s t_{FM} t_{HM}}). \quad (M14)$$

When $S_{SP+heat}$ and thus $\xi_{FMR,SP+heat}$ are negligible, $1/\xi_{FMR}$ is the same as $1/\xi_{FMR,y}$ and a linear function of $1/t_{FM}$ (as is the case of the Ta (5)/Py(< 4.6), the Ir (5)/Py (4.6-7.6), and the Pt (4)/Py (3.3-8.6) in this work, Fig. 6b). A linear fit of from $1/\xi_{FMR}$ vs $1/t_{FM}$ can be used to determine $\xi_{DL,y}$ (the inverse intercept) and $\xi_{FL,y}$ (from the slope). From the linear scaling, $\xi_{FMR,y}$ and $S_{DL,y}$ can be extrapolated for any $t_{FM}$. When $S_{SP+heat}$ and thus $\xi_{FMR,SP+heat}$ become nonzero, $1/\xi_{FMR}$ vs $1/t_{FM}$ deviates from the linear scaling given by $1/\xi_{FMR,y}$ vs $1/t_{FM}$ (as is the case of the Ta (5)/Py($\geq$ 4.6) in this work, Fig. 6b). From the deviation from the linear scaling extrapolated from the thin-limit samples Ta (5)/Py(< 4.6), $\xi_{FMR,SP+heat}$ and thus $S_{SP+heat}$ (Fig. 6c) can be determined for the thick-limit samples Ta (5)/Py(< 4.6). Therefore, a linear dependence of $1/\xi_{FMR}$ on $1/t_{FM}$ always signifies the absence of any important $S_{SP+heat}$. We also stress that the presence of any $S_{SP+heat}$ will not affect the signals of $z$ spins ($S_{DL,z}$ and $A_{DL,z}$) since they are separated from the $\varphi$ dependence of the symmetric ST-FMR signal following Equation (1). Note that the determined $\xi^j_{DL,z}$ values are essentially independent of the rf power and thus thermoelectric effects during the ST-FMR measurement, within the experimental uncertainty (Fig. 6d).

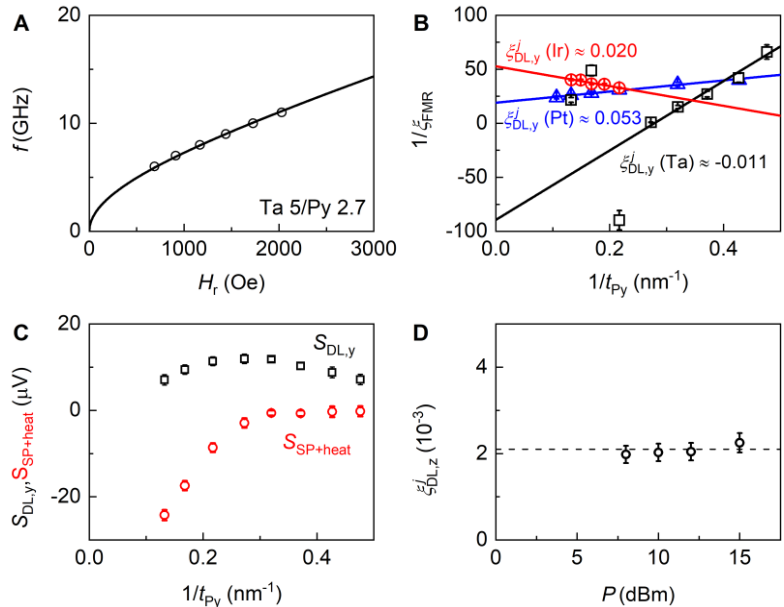

**Fig. 6| More ST-FMR data.** (**a**) Ferromagnetic resonance field ($H_r$) vs the rf frequency ($f$) of the Ta (5)/Py (2.7). The solid line represents the best fit to Kittel's equation. (**b**) $1/\xi_{FMR}$ vs $1/t_{Py}$ for the Ta/Py, the Ir/Py, and the Pt/Py as measured using 3-terminal ST-FMR. From the linear fits, $\xi^j_{DL,y}$ is determined to be -0.011 ± 0.001 for the Ta/Py, 0.020± 0.001 for the Ir/Py, and 0.053 ± 0.003 for the Pt/Py, respectively. (**c**) The values of $S_{DL,y}$ and $S_{SP+heat}$ vs $1/t_{Py}$ for Ta (5)/Py ($t_{Py}$) devices as separated from the deviation of the linear scaling

predicted by the thin-limit data in (b). The non-monotonic dependence of $S_{DL,y}$ is attributed to the opposite thickness dependences of the magnetoresistance ratio ($C_{MR}$) and dampinglike torque field of the Py layer. (**d**) The values of $\xi_{DL,z}^j$ for the Ta (5)/Py (2.7) measured from the 2-terminal ST-FMR using different rf power.

**References**

[1] W. W. Lin, H. Sang, D. Liu, Z. S. Jiang, A. Hu, X. S. Wu, Magnetization switching induced by in-plane current with low density in Pt/Co/Pt Sandwich, J. Appl. Phys. **99**, 08G518 (2006).

[2] I. M. Miron, K. Garello, G. Gaudin, P.-J. Zermatten, M. V. Costache, S. Auffret, S. Bandiera, B. Rodmacq, A. Schuhl, P. Gambardella, Perpendicular switching of a single ferromagnetic layer induced by in-plane current injection, Nature (London) **476**, 189 (2011).

[3] S. Fukami, C. Zhang, S. DuttaGupta, A. Kurenkov, H. Ohno, Magnetization switching by spin-orbit torque in an antiferromagnet-ferromagnet bilayer system, Nat. Mater. **15**, 535 (2016).

[4] Y.-W. Oh, S.C. Baek, Y. M. Kim, H.Y. Lee, K.-D. Lee, C.-G. Yang, E.-S. Park, K.-S. Lee, K.-W. Kim, G. Go, J.-R. Jeong, B.-C. Min, H.-W. Lee, K.-J. Lee, B.-G. Park, Field-free switching of perpendicular magnetization through spin-orbit torque in antiferromagnet/ferromagnet/oxide structures, Nat. Nanotech. **11**, 878 (2016).

[5] Y.-C. Lau, D. Betto, K. Rode, J. M. D. Coey, P. Stamenov, Spin-orbit torque switching without an external field using interlayer exchange coupling, Nat. Nanotechnol. **11**, 758 (2016).

[6] G. Yu, P. Upadhyaya, Y. Fan, J. G. Alzate, W. Jiang, K. L. Wong, S. Takei, S. A. Bender, L.-T. Chang, Y. Jiang, M. Lang, J. Tang, Y. Wang, Y. Tserkovnyak, P.K. Amiri, K. L. Wang, Switching of perpendicular magnetization by spin-orbit torques in the absence of external magnetic fields, Nat. Nanotech. **9**, 548 (2014).

[7] K. Cai, M. Yang, H. Ju, S. Wang, Y. Ji, B. Li, K. W. Edmonds, Y. Sheng, B. Zhang, N. Zhang, S. Liu, H. Zheng, Electric field control of deterministic current-induced magnetization switching in a hybrid ferromagnetic/ferroelectric structure, Nat. Mater. **16**, 712 (2017).

[8] M. Wang, W. Cai, D. Zhu, Z. Wang, J. Kan, Z. Zhao, K. Cao, Z. Wang, Y. Zhang, T. Zhang, C. Park, J.-P. Wang, A. Fert, W. Zhao, Field-free switching of a perpendicular magnetic tunnel junction through the interplay of spin-orbit and spin-transfer torques. Nat. Electron. **1**, 582 (2018).

[9] N. Sato, F. Xue, R. M. White, C. Bi, S. X. Wang, Two-terminal spin-orbit torque magnetoresistive random access memory, Nat. Electron. **1**, 508 (2018).

[10] O. J. Lee, L. Q. Liu, C. F. Pai, Y. Li, H. W. Tseng, P. G. Gowtham, J. P. Park, D. C. Ralph, R. A. Buhrman, Central role of domain wall depinning for perpendicular magnetization switching driven by spin torque from the spin Hall effect, Phys. Rev. B **89**, 024418 (2014).

[11] L. Zhu, Switching of Perpendicular Magnetization by Spin-Orbit Torque, Adv. Mater. **35**, 2300853 (2023).

[12] D. MacNeill, G. M. Stiehl, M. H. D. Guimaraes, R. A. Buhrman, J. Park, D. C. Ralph, Control of spin-orbit torques through crystal symmetry in WTe$_2$/ferromagnet bilayers, Nat. Phys. **13**, 300 (2017).

[13] A. M. Humphries, T. Wang, E. R. J. Edwards, S. R. Allen, J. M. Shaw, H. T. Nembach, J. Q. Xiao, T. J. Silva, X. Fan, Observation of spin-orbit effects with spin rotation symmetry, Nat. Commun. **8**, 911 (2017).

[14] S. C. Baek, V. P. Amin, Y. Oh, G. Go, S. Lee, G. Lee, K. Kim, M. D. Stiles, B. Park, K. Lee, Spin currents and spin-orbit torques in ferromagnetic trilayers, Nat. Mater. **17**, 509 (2018).

[15] L. Liu, C. Zhou, X. Shu, C. Li, T. Zhao, W. Lin, J. Deng, Q. Xie, S. Chen, J. Zhou, R. Guo, H. Wang, J. Yu, S. Shi, P. Yang, S. Pennycook, A. Manchon, J. Chen, Symmetry-dependent field-free switching of perpendicular magnetization, Nat. Nanotech. **16**, 277 (2021).

[16] T. Nan, C. X. Quintela, J. Irwin, G. Gurung, D. F. Shao, J. Gibbons, N. Campbell, K. Song, S.-Y. Choi, L. Guo, R. D. Johnson, P. Manuel, R. V. Chopdekar, I. Hallsteinsen, T. Tybell, P. J. Ryan, J.-W. Kim, Y. Choi, P. G. Radaelli, D. C. Ralph, E. Y. Tsymbal, M. S. Rzchowski, C. B. Eom, Controlling spin current polarization through non-collinear antiferromagnetism, Nat. Commun. **11**, 4671 (2020).

[17] Y. You, H. Bai, X. Feng, X. Fan, L. Han, X. Zhou, Y. Zhou, R. Zhang, T. Chen, F. Pan, C. Song, Cluster magnetic octupole induced out-of-plane spin polarization in antiperovskite antiferromagnet, Nat. Commun. **12**, 6524 (2021).

[18] A. Bose, N. J. Schreiber, R. Jain, D. Shao, H. P. Nair, J. Sun, X. S. Zhang, D. A. Muller, E. Y. Tsymbal, D. G. Schlom, D. C. Ralph, Tilted spin current generated by the collinear antiferromagnet ruthenium dioxide, Nat. Electron. **5**, 267 (2022).

[19] H. Bai, Y. C. Zhang, Y. J. Zhou, P. Chen, C. H. Wan, L. Han, W. X. Zhu, S. X. Liang, Y. C. Su, X. F. Han,

F. Pan, C. Song, Efficient Spin-to-Charge Conversion via Altermagnetic Spin Splitting Effect in Antiferromagnet $RuO_2$, Phys. Rev. Lett. **130**, 216701 (2022).

[20] S. Karube, T. Tanaka, D. Sugawara, N. Kadoguchi, M. Kohda, J. Nitta, Observation of Spin-Splitter Torque in Collinear Antiferromagnetic $RuO_2$, Phys. Rev. Lett. **129**, 137201(2022).

[21] S. Hu, D. Shao, H. Yang, C. Pan, Z. Fu, M. Tang, Y. Yang, W. Fan, S. Zhou, E. Y. Tsymbal, X. Qiu, Efficient perpendicular magnetization switching by a magnetic spin Hall effect in a noncollinear antiferromagnet, Nat. Commun. **13**, 4447 (2022).

[22] M. DC, D. Shao, V. D.-H. Hou, P. Quarterman, A. Habiboglu, B. Venuti, M. Miura, B. Kirby, A. Vailionis, C. Bi, X. Li, F. Xue, Y. Huang, Y. Deng, S. Lin, W. Tsai, S. Eley, W. Wang, J. A. Borchers, E. Y. Tsymbal, S. X. Wang, Observation of anti-damping spin–orbit torques generated by in-plane and out-of-plane spin polarizations in $MnPd_3$, Nat. Mater. **22**, 591 (2023).

[23] D. Meng, S. Chen, C. Ren, J. Li, G. Lan, C. Li, Y. Liu, Y. Su, G. Yu, G. Chai, R. Xiong, W. Zhao, G. Yang, S. Liang, Field-Free Spin-Orbit Torque Driven Perpendicular Magnetization Switching of Ferrimagnetic Layer Based on Noncollinear Antiferromagnetic Spin Source, Adv. Electron. Mater. 10, 2300665 (2023).

[24] A. Davidson, V. P. Amin, W. S. Aljuaid, P. M. Haney, X. Fan, Perspectives of electrically generated spin currents in ferromagnetic materials, Phys. Lett. A **384**, 126228 (2020).

[25] L. Zhu, X. S. Zhang, D. A. Muller, D. C. Ralph, R. A. Buhrman, Observation of strong bulk damping-like spin-orbit torque in chemically disordered ferromagnetic single layers, Adv. Funct. Mater. **30**. 2005201 (2020).

[26] L. Liu, T. Moriyama, D. C. Ralph, R. A. Buhrman, Spin-torque ferromagnetic resonance induced by the spin Hall effect, Phys. Rev. Lett. **106**, 036601 (2011).

[27] C. Pai, Y. Ou, L. H. Vilela-Leão, D. C. Ralph, R. A. Buhrman, Dependence of the efficiency of spin Hall torque on the transparency of Pt/ferromagnetic layer interfaces, Phys. Rev. B **92**, 064426 (2015).

[28] O. Mosendz, V. Vlaminck, J. E. Pearson, F. Y. Fradin, G. E. W. Bauer, S. D. Bader, A. Hoffmann, Detection and quantification of inverse spin Hall effect from spin pumping in permalloy/normal metal bilayers, Phys. Rev. B **82**, 214403 (2010).

[29] S. Karimeddiny, J. A. Mittelstaedt, R. A. Buhrman, and D. C. Ralph, Transverse and Longitudinal Spin-Torque Ferromagnetic Resonance for Improved Measurement of Spin-Orbit Torque, Phys. Rev. Applied 14, 024024 (2020).

[30] Q. Liu, L. Zhu, Current-induced perpendicular effective magnetic field in magnetic heterostructures, Appl. Phys. Rev. **9**, 041401 (2022).

[31] M. Harder, Y. Gui, C.-M. Hu, Electrical detection of magnetization dynamics via spin rectification effects, Phys. Rep. **661**,1 (2016).

[32] D. Fang, H. Kurebayashi, J. Wunderlich, K. Výborný, L. P. Zârbo, R. P. Campion, A. Casiraghi, B. L. Gallagher, T. Jungwirth, A. J. Ferguson, Spin-orbit-driven ferromagnetic resonance, Nature Nanotech. **6**, 413 (2011).

[33] T.-Y. Chen, Y. Ou, T.-Y. Tsai, R. A. Buhrman, C.-F. Pai, Spin-orbit torques acting upon a perpendicularly magnetized Py layer, APL Mater. **6**, 121101 (2018).

[34] K. Kondou, H. Sukegawa, S. Kasai, S. Mitani, Y. Niimi, Y. Otani, Influence of inverse spin Hall effect in spin-torque ferromagnetic resonance measurements, Appl. Phys. Express **9**, 023002 (2016).

[35] S. Dutta, A. A. Tulapurkar, Observation of nonlocal orbital transport and sign reversal of dampinglike torque in Nb/Ni and Ta/Ni bilayers, Phys. Rev. B **106**, 184406 (2022).

[36] T. Tanaka, H. Kontani, M. Naito, T. Naito, D. S. Hirashima, K. Yamada, and J. Inoue, Intrinsic spin Hall effect and orbital Hall effect in 4d and 5d transition metals, Phys. Rev. B **77**, 165117 (2008).

[37] D. Lee, D. Go, H. Park, W. Jeong, H. Ko, D. Yun, D. Jo, S. Lee, G. Go, J. H. Oh, K. Kim, B. Park, B. Min, H. C. Koo, H. Lee, O. Lee, K. Lee, Orbital torque in magnetic bilayers, Nat. Commun. **12**, 6710 (2021).

[38] C. O. Avci, K. Garello, M. Gabureac, A. Ghosh, A. Fuhrer, S. F. Alvarado, P. Gambardella, Interplay of spin-orbit torque and thermoelectric effects in ferromagnet/normal-metal bilayers, Phys. Rev. B **90**, 224427 (2014).

[39] D. MacNeill, G. M. Stiehl, M. H. D. Guimarães, N. D. Reynolds, R. A. Buhrman, D. C. Ralph, Thickness dependence of spin-orbit torques generated by $WTe_2$, Phys. Rev. B **96**, 054450 (2017).

[40] M. G. Kang, J.-G. Choi, J. Jeong, J. Y. Park, H.-J. Park, T. Kim, T. Lee, K.-J. Kim, K.-W. Kim, J. H. Oh, D. D. Viet, J.-R. Jeong, J. M. Yuk, J. Park, K.-J. Lee, B.-G. Park, Electric-field control of field-free spin-orbit torque switching via laterally modulated Rashba effect in Pt/Co/$AlO_x$ structures, Nat. Commun. **12**, 7111 (2021).

[41] S. L. Yin, Q. Mao, Q. Y. Meng, D. Li, H. W. Zhao, Hybrid anomalous and planar Nernst effect in

permalloy thin films, Phys. Rev. B **88**, 064410 (2013).

[42] A. Ghosh, K. Garello, C. O. Avci, M. Gabureac, P. Gambardella, Interface-Enhanced Spin-Orbit Torques and Current-Induced Magnetization Switching of Pd/Co/AlO$_x$ Layers, Phys. Rev. Appl. **7**, 014004 (2017).

[43] Q. Liu, X. Lin, L. Zhu, Absence of Spin-Orbit Torque and Discovery of Anisotropic Planar Nernst Effect in CoFe Single Crystal, Adv. Sci. **10**, 2301409 (2023).

[44] F. Xue and P. M. Haney, Staggered spin Hall conductivity, Phys. Rev. B **102**, 195146 (2020).

[45] A. Roy, M. H. D. Guimarães, and J. Sławińska, Unconventional spin Hall effects in nonmagnetic solids, Phys. Rev. Materials **6**, 045004 (2022).

[46] H. B. M. Saidaoui and A. Manchon, Spin-Swapping Transport and Torques in Ultrathin Magnetic Bilayers, Phys. Rev. Lett. **117**, 036601 (2016).

[47] L. Zhu, D. C. Ralph, R. A. Buhrman, Spin-orbit torques in heavy-metal-ferromagnet bilayers with varying strengths of interfacial spin-orbit coupling, Phys. Rev. Lett. **122**, 077201 (2019).

[48] Q. Liu, L. Liu, G. Xing, L. Zhu, Asymmetric magnetization switching and programmable complete Boolean logic enabled by long-range intralayer Dzyaloshinskii-Moriya interaction, Nat. Commun. 15, 2978 (2024).

[49] A. Fernández-Pacheco, E. Vedmedenko, F. Ummelen, R. Mansell, D. Petit, R. P. Cowburn, Symmetry-breaking interlayer Dzyaloshinskii–Moriya interactions in synthetic antiferromagnets, Nat. Mater. 18, 679–684 (2019).

[50] J. Moritz, F. Garcia, J. C. Toussaint, B. Dieny, and J. P. Nozi`eres, Orange peel coupling in multilayers with perpendicular magnetic anisotropy: Application to (Co=Pt)-based exchange-biased spin-valves, Europhys. Lett. 65, 123 (2004).

[51] Y. Guo, Y. Wu, Y. Cao, X. Zeng, B. Wang, D. Yang, X. Fan, and J. Cao, The deterministic field-free magnetization switching of perpendicular ferrimagnetic Tb-Co alloy film induced by interfacial spin current, Appl. Phys. Lett. 119, 032409 (2021).

[52] Q. Yang, D. Han, S. Zhao, J. Kang, F. Wang, S. Lee, J. Lei, K. Lee, B. Park, and H. Yang, Field-free spin-orbit torque switching in ferromagnetic trilayers at sub-ns timescales. Nat. Commun. **15**, 1814 (2024).

[53] Y. Cao, Y. Sheng, K. W. Edmonds, Y. Ji, H. Zheng, and K. Wang, Deterministic Magnetization Switching Using Lateral Spin-Orbit Torque. Adv. Mater. 32, 1907929 (2020).

[54] K. Hasegawa, T. Koyama, and D. Chiba, Current-induced perpendicular magnetization switching without external magnetic field in gate-induced asymmetric structure. Appl. Phys. Lett. 119, 202402 (2021).

[55] J. Ryu, R. Thompson, J. Y. Park, *et al.* Efficient spin–orbit torque in magnetic trilayers using all three polarizations of a spin current. Nat. Electron. **5**, 217-223 (2022).

[56] T. Jin, G. J. Lim, H. Y. Poh, S. Wu, F. Tan, and W. S. Lew, Spin Reflection-Induced Field-Free Magnetization Switching in Perpendicularly Magnetized MgO/Pt/Co Heterostructures, ACS Appl. Mater. Interfaces 14, 7 (2022).

[57] L. Liu, C. Zhou, T. Zhao, *et al.* Current-induced self-switching of perpendicular magnetization in CoPt single layer. Nat Commun **13**, 3539 (2022).

[58] J. Li, Q. Guo, T. Lin, Q. Zhang, H. Bai, S. Cheng, X. Zhan, L. Gu, and T. Zhu, Interface Effect on the Out-of-Plane Spin-Orbit Torque in the Ferromagnetic CoPt Single Layers. Adv. Funct. Mater. 2401018 (2024).

[59] W. Wang, Q. Fu, K. Zhou, L. Chen, L. Yang, Z. Li, Z. Tao, C. Yan, L. Liang, X. Zhan, Y. Du, and R. Liu, Unconventional Spin Currents Generated by the Spin-Orbit Precession Effect in Perpendicularly Magnetized Co-Tb Ferrimagnetic System. Phys. Rev. Appl. 17, 034026 (2022).

[60] X. Shu, J. Zhou, L. Liu, W. Lin, C. Zhou, S. Chen, Q. Xie, L. Ren, Y. Xiaojiang, H. Yang, and J. Chen, Role of Interfacial Orbital Hybridization in Spin-Orbit-Torque Generation in Pt-Based Heterostructures. Phys. Rev. Appl. 14, 054056 (2020).

[61] Z. Gong, F. Liu, X. Guo and C. Jiang, Observation of the out-of-plane orbital antidamping-like torque, Phys. Chem. Chem. Phys. 26, 6345-6350 (2024).

[62] F. Xue, S. J. Lin, M. Song, W. Hwang, C. Klewe, C.-M. Lee, E. Turgut, P. Shafer, A. Vailionis, Y.-L. Huang, W. Tsai, X. Bao, S. X. Wang, Field-free spin-orbit torque switching assisted by in-plane unconventional spin torque in ultrathin [Pt/Co]$_N$. Nat Commun. 14, 3932 (2023).

[63] B. K. Hazra, B. Pal, J. Jeon, R. R. Neumann, B. Göbel, B. Grover, H. Deniz, A. Styervoyedov, H. Meyerheim, I. Mertig, S. Yang, S. S. P. Parkin, Generation of out-of-plane polarized spin current by spin swapping, Nat. Commun. **14**, 4549 (2023).

[64] P. Gupta, N. Chowdhury, M. Xu, P. K. Muduli, A. Kumar, K. Kondou, Y. Otani, and P. K. Muduli, Generation of out-of-plane polarized spin current in (permalloy, Cu)/EuS interfaces, Phys. Rev. B 109,

L060405 (2024).

[65] Q. Fu, L. Liang , W. Wang, L. Yang, K. Zhou, Z. Li, C. Yan, L. Li, H. Li, and R. Liu. Observation of nontrivial spin-orbit torque in single-layer ferromagnetic metals. Phys. Rev. B 105, 224417 (2022).

[66] D. Tiwari, N. Behera, A. Kumar, P. Dürrenfeld, S. Chaudhary, D. K. Pandya, J. Åkerman, P. K. Muduli. Appl. Phys. Lett. 111, 232407 (2017).

[67] V. P. Amin, J. Zemen, and M. D. Stiles, Interface-Generated Spin Currents, Phys. Rev. Lett. 121, 136805 (2018).

[68] L. Zhu and R. A. Buhrman, Absence of Significant Spin-Current Generation in Ti/Fe−Co−B Bilayers with Strong Interfacial Spin-Orbit Coupling, Phys. Rev. Applied 15, L031001 (2021).

[69] L. Zhu, D.C. Ralph, R.A. Buhrman, Lack of Simple Correlation between Switching Current Density and Spin-Orbit-Torque Efficiency of Perpendicularly Magnetized Spin-Current-Generator-Ferromagnet Heterostructures, Phys. Rev. Appl. **15**, 024059 (2021).

[70] J.-W. Xu and A. D. Kent, Charge-To-Spin Conversion Efficiency in Ferromagnetic Nanowires by Spin Torque Ferromagnetic Resonance: Reconciling Lineshape and Linewidth Analysis Methods, Phys. Rev. Applied 14, 014012 (2020).

[71] V. Kateel, Viola. Krizakova, S. Rao, K. Cai, M. Gupta, M. G. Monteiro, F. Yasin, B. Sorée, J. De Boeck, S. Couet, P. Gambardella, G. S. Kar, and K. Garello, Field-Free Spin–Orbit Torque Driven Switching of Perpendicular Magnetic Tunnel Junction through Bending Current, Nano Lett. **23**, 5482 (2023).

[72] M. Wang, J. Zhou, X. Xu, T. Zhang, Z. Zhu, Z. Guo, Y. Deng, M. Yang, K. Meng, B. He, J. Li, G. Yu, T. Zhu, A. Li, X. Han, Y. Jiang, Field-free spin-orbit torque switching via out-of-plane spin-polarization induced by an antiferromagnetic insulator/heavy metal interface, Nat. Commun. **14**, 2871 (2023).

**Author contribution**
L.Z. conceived the project, Q. L. fabricated the Py samples and performed majority of the measurements, X. L. performed the finite-element analysis, A.S. performed part of the ST-FMR measurements under construction of Q. L., Z.N. and G.Y sputter-deposited the PMA Ta/FeCoB sample. L. Z. and Q. L. wrote the manuscript, all the authors discussed the results and contributed to the manuscript writing.

**Acknowledgements**
The authors thank Daniel C Ralph for critical reading of the manuscript and Wanjun Jiang, Jiahao Liu, Jingrun Chen and Yifei Sun for discussions on simulation of the electric field distributions. This work was supported partly by the National Key Research and Development Program of China (2022YFA1204000), by the Beijing National Natural Science Foundation (Z230006), by the Strategic Priority Research Program of the Chinese Academy of Sciences (XDB44000000), by the National Natural Science Foundation of China (12304155,12274405). Measurements at Cornell were supported by the DOE under award number DE-SC0017671.

**Conflict of interests:** The authors declare no conflict of interests.

**Data Availability:** The data that support the findings of this study are available from the corresponding author upon reasonable request.

**Keywords:** spin current, spin-orbit torque, magnetization switching